\documentclass[aps,prb,twocolumn,superscriptaddress,english]{revtex4-2}
\usepackage[utf8]{inputenc}

\usepackage{amsmath}
\usepackage{amssymb}
\usepackage{graphicx}
\usepackage{siunitx}
\usepackage{lipsum}
\usepackage[table]{xcolor}
\usepackage{rotating}
\usepackage{placeins}
\usepackage{lineno}

\newcommand{\tc}[1]{$T_{c}$}
\newcommand{\ep}[1]{$ep$}

\usepackage{hyperref}
\usepackage{orcidlink}
\usepackage{xr-hyper}

\definecolor{softred}{rgb}{0.92, 0.80, 0.80}
\definecolor{softgreen}{rgb}{0.82, 0.90, 0.82}

\begin{document}

\author{Simone Di Cataldo  \orcidlink{0000-0002-8902-0125}}\email{simone.dicataldo@uniroma1.it}
\affiliation{Dipartimento di Fisica, Sapienza Universit\`a di Roma, Piazzale Aldo Moro 5, 00187 Roma, Italy}
\author{William Cursio}
\affiliation{Dipartimento di Fisica, Sapienza Universit\`a di Roma, Piazzale Aldo Moro 5, 00187 Roma, Italy}
\affiliation{Leibniz Institute for Solid State and Materials Science Dresden (IFW Dresden), 01069 Dresden, Germany}
\author{Lilia Boeri \orcidlink{0000-0003-1186-2207}}
\affiliation{Dipartimento di Fisica, Sapienza Universit\`a di Roma, Piazzale Aldo Moro 5, 00187 Roma, Italy}

\title{Vacancy-Controlled Superconductivity in Rock-Salt Carbides: Towards Predictive Modelling of Real-World Superconductors}

\begin{abstract}
We critically reexamine the superconducting properties of rock-salt transition-metal carbides (TMCs), often regarded as textbook conventional superconductors, combining first-principles electron–phonon calculations with variable-composition evolutionary structure prediction. Studying superconducting trends across the entire transition-metal series, we find that, when the rock-salt stoichiometric phase is dynamically or thermodynamically unstable, carbon-vacant structures identified through unbiased structure prediction permit to reconcile theoretical calculations with experimental trends. Our integrated use of structure prediction and electron–phonon calculations defines a general framework for realistic modelling of superconductors shaped by non-equilibrium synthesis routes and defect tolerance.
\end{abstract}

\maketitle
\section{Introduction}
\begin{table*}[!htb] 
    \centering
    \begin{tabular}{|c|c|c|c|c|c|c|c||c|c|c|c|c|c|}
    \hline
    Comp. & Exp. \tc{} & Dyn. & $\Delta H$   & N(E$_F$) & T$_c^{McM}$ & $\lambda$ & $\omega_{log}$ & Alt. & Dyn. & N(E$_F$)  & T$_c^{McM}$ &$\lambda$ & $\omega_{log}$ \\
            &     (K)    &    st.    &  (meV/at.) &(eV$\cdot$sp.$\cdot$at.)$^{-1}$&    (K)      &           &    (K)     &  Comp.   &   st.   & (eV$\cdot$sp.$\cdot$at.)$^{-1}$&  (K)  &           &    (K) \\
    \hline
    \rowcolor{softred} ScC  & $<$ 1.4 \cite{Nowotny_1961_ScC, Samsonov_1962_ScC} &Y& 106 
    & 0.36 & 5 & 0.6 & 461 & Sc$_{6}$C$_{5}$  & Y& 0.37 & 2 &   0.5   & 417 \\
    \rowcolor{softred} YC   & $<$ 1.4 &Y& 356 
    & 0.41 & 10 & 0.8 & 364 & Y$_{6}$C$_5$ &Y& 0.36 & 5 &  0.6 &  328 \\
\hline
    \rowcolor{softgreen} TiC  & 0--3.4 \cite{Klimashin_1970_TiC, Toth_2014_TMC} &Y& 0 
    & 0.07 & 0 & 0.0 & 562 & Ti$_{6}$C$_{5}$ &Y& 0.15 & 0 &  0.2  &  559\\
    \rowcolor{softgreen} ZrC  & $<$ 0.3 &Y& 0 
    & 0.06 & 0 & 0.2 & 473 & Zr$_{6}$C$_{5}$ &Y& 0.13 & 0 & 0.3 & 455 \\
    \rowcolor{softgreen} HfC  & $<$ 1.2 &Y& 0 
    & 0.07 & 0 & 0.2 & 427 & Hf$_{6}$C$_{5}$ &Y& 0.12 & 0 & 0.3 & 394 \\
\hline
    \rowcolor{softred} VC   & 0--3.2 \cite{Geerk_1977_implantation_VC} &Y& 92 
    & 0.28 & 17 & 1.0 & 309 & V$_6$C$_5$ &Y& 0.31 & 0 &   0.0  &   483   \\
    \rowcolor{softgreen} NbC  & 0--11.5 \cite{Giorgi_PR_1962_vacancy, Willens_PR_1967_TMC_SC, Toth_ActMet_1968_NbC_vac, Yan_PRB_2020_NbC_arpes_sc, Shang_PRB_2020_NbC_TaC} &Y& 28 
    & 0.19 & 15 & 1.0 & 299 & Nb$_{6}$C$_{5}$ &Y& 0.17 & 1 & 0.4  & 389 \\
    \rowcolor{softgreen} TaC  & 0--10.3 \cite{Giorgi_PR_1962_vacancy, Toth_ActMet_1968_NbC_vac, Shang_PRB_2020_NbC_TaC} &Y& 0 
    & 0.16 & 7 & 0.7 & 259 & Ta$_{6}$C$_{5}$ &Y& 0.17 & 1 & 0.5  & 301 \\
\hline
    \rowcolor{softred} MoC  & 14.1 \cite{Clougherty_1961_MoC, Willens_AppPhysLett_1965_WC_MoC, Sadagapan_1966_MoC, Toth_1968_MoC, Willens_PR_1967_TMC_SC, Sathish_2012_MoC068} &N& 288 
    & 0.33 & - & - & - & Mo$_{6}$C$_{5}$ &Y& 0.32 & 18  &  1.4  & 209 \\
    \rowcolor{softred} WC   & 3.5--10 \cite{Willens_AppPhysLett_1965_WC_MoC, Willens_PR_1967_TMC_SC} &N& 420 
    & 0.28 & - & - & - & W$_{6}$C$_{5}$ &Y& 0.16 &  5 & 0.7 & 239 \\
\hline
    \rowcolor{softred} ReC  & 3.4 \cite{Popova_1972_ReC} & N & 932 
    & 0.59 & - & - & - & Re$_6$C$_5$ & N & 0.3 & - &  -  & - \\
    \hline
    \end{tabular}
    \caption{Summary of the thermodynamic and superconducting properties of various TMCs in the stoichiometric and vacant rock-salt structure. Green (red) cells highlight thermodynamical stability (instability) of the stoichiometric rock-salt phase. In red cells the calculated \tc{} for the stoichiometric phase is also in disagreement with experiment. $\Delta H$, $N(E_F)$, $T_{C}^{McM}$, $\lambda$, and $\omega_{log}$ indicate the energy relative to the convex hull, the DOS at the Fermi energy, the calculated \tc{}, the \ep{} coupling coefficient, and the average phonon frequency, respectively. Columns 1-8 indicate the properties of the stoichiometric (1:1) phase, while columns 9-14 indicate those of carbon-vacant (6:5) phase, for which the calculated and measured \tc{} values are in agreement. When available, direct references to experimental papers are given in the table. Other values are taken from Refs. \cite{Matthias_RevModPhys_1963, Vonsovsky_1982_TMS, Toth_2014_TMC}. The \tc{}'s are computed using the McMillan formula \cite{McMillanTC, Allen_PRB_1975_transition} with $\mu^{*} = 0.15$.
    }
    \label{tab:tcs_table}
\end{table*}

Transition metal carbides (TMCs) are a broad class of materials formed by  early transition metals and carbon, which crystallise in a rock-salt (NaCl-type) structure ~\cite{Williams_1971_review}. They serve as high-performance materials for cutting-tools, coatings, heat and chemical shields, due to their exceptional resistance. These carbides are metallic conductors ($\rho \sim$ 10$^{-5} \Omega m$) \cite{Modine_1989_electricaltmc} with high thermal conductivity \cite{Radosevich_1970_thermaltmc}. When the transition metal (TM) belongs to groups V (V, Nb, Ta) and VI (Mo, W), TMCs  also exhibit superconductivity, with critical temperatures (\tc{}) reaching up to 18~K in NbN-NbC-TiN alloys~\cite{Toth_JPCS_1965_MoC, Toth_1968_MoC, Pessall_JPCS_1968_TMC_alloy, Morton_JCM_1971_MoC_WC, Vonsovsky_1982_TMS, Toth_2014_TMC}, and high critical fields \cite{Fink_1965_TMCfield}. 

So far, TMCs have found limited use in superconducting technologies due to their brittleness.
However, there is a clear potential for applications as shock- and radiation-hard superconductors in extreme environments - such as cryogenic current leads on deep-space probes, kinetic-energy dampers, or compact fusion diagnostics - where the unparalleled hardness, erosion resistance and thermal stability of TMCs could outweigh manufacturing challenges.
Understanding in detail the relationships between crystal chemistry and superconducting properties
is a crucial prerequisite to further optimize these materials and unlock their 
full technological potential.

To date, experimental data remains sparse and largely outdated,
with most superconducting critical temperatures (\tc{}) reported in the 1960s and 1970s
from samples synthesized at high temperatures.  
The data for TMCs with a finite \tc{} is summarized in Tab. \ref{tab:tcs_table},
where we included, to the best of our knowledge, all experimental reports \cite{Nowotny_1961_ScC, Samsonov_1962_ScC, Klimashin_1970_TiC, Toth_2014_TMC,
Geerk_1977_implantation_VC,Giorgi_PR_1962_vacancy, Willens_PR_1967_TMC_SC, Toth_ActMet_1968_NbC_vac, Yan_PRB_2020_NbC_arpes_sc, Shang_PRB_2020_NbC_TaC,
Clougherty_1961_MoC, Willens_AppPhysLett_1965_WC_MoC, Sadagapan_1966_MoC, Toth_1968_MoC, Willens_PR_1967_TMC_SC, Sathish_2012_MoC068,Popova_1972_ReC,Matthias_RevModPhys_1963, Vonsovsky_1982_TMS, Toth_2014_TMC}.
The table highlights a high variability in reported \tc{}'s
for elements in groups V and VI, which has been historically attributed
to differences in carbon vacancy concentrations.
 \cite{Giorgi_PR_1962_vacancy, Toth_ActMet_1968_NbC_vac, Schwarz_1976_NbC_vacancies, Geerk_1977_implantation_VC, Pickett_1986_NbC_vac}.

The microscopic understanding of superconductivity in TMCs is still superficial and fails to explain most experimental observations. In particular, although \tc{} is known to depend strongly on carbon content, existing studies on the role of vacancies are limited to the thermodynamic and mechanical stability of nonstoichiometric phases~\cite{Jang_ActaM_2012_TiC_vacancies, Muchiri_CMS_2022_NbC_NbC_vac}. All first-principles studies of superconductivity so far have been based on the ideal 1:1 stoichiometric rock-salt structure, implicitly assuming its validity across the entire composition range. Recent work on group-IV and group-V compounds has interpreted variations in \tc{} in terms of the effect of electron filling of the transition-metal $d$ states on the electron–phonon interaction~\cite{Shang_PRB_2020_NbC_TaC, Noffsinger_2008_HfCTaC, Sun_PRB_2015_TiC_vac}, but failed to address the strong dependence of \tc{} on vacancy concentration highlighted in Table~\ref{tab:tcs_table}. TMCs containing group-VI elements were shown to be dynamically unstable in the 1:1 rock-salt structure, highlighting even further the limitations of describing these compounds in the stoichiometric phase~\cite{Isaev_JAP_2007_TMCs_tc, Connetable_2016_TMC}.


In this work, we go beyond the 1:1 stoichiometric limit and systematically investigate the role of carbon vacancies on superconductivity across the TMC series, combining \textit{ab initio} electron-phonon superconductivity theory and crystal structure prediction.
Using unbiased evolutionary searches at variable compositions, we construct the TM-C phase diagrams, explicitely addressing the effect of non-stoichiometry.
This leads us to identify a family of low-energy nonstoichiometric phases that preserve the rock-salt structure type, providing a natural explanation for the superconducting properties observed experimentally across the whole TMC series considered.

The main results across the whole TMC series are summarized by color-coding in table  \ref{tab:tcs_table}. The TMs are ordered by the groups of the periodic table, and colored in green when the stoichiometric phase is thermodynamically (meta)stable (within 50 meV/atom from the hull), and red otherwise. The table reports the main superconducting parameters for the stoichiometric (1:1) phase, and a carbon-vacant (6:5) phase which preserves the rock-salt geometry. We note that elements of group IV and V (Nb and Ta) are the only ones for which the stoichiometric phase is stable and here the calculated and measured values of \tc{} match. In groups III, VI and VII, the stoichiometric phase is instead unstable, and the calculated \tc{}'s do not match experiments, while the carbon-vacant phase is found in much better agreement.

Our results demonstrate that thermodynamic stability is a crucial element in the realistic modeling of real-world superconductors.\cite{Schwarz_1976_NbC_vacancies, Klein_1980_CPA_ep}.


\section{Results and discussion}
\subsection{Thermodynamical properties}
\begin{turnpage}
\begin{figure*}[p]
    \centering
    \includegraphics[width=2.80\columnwidth]{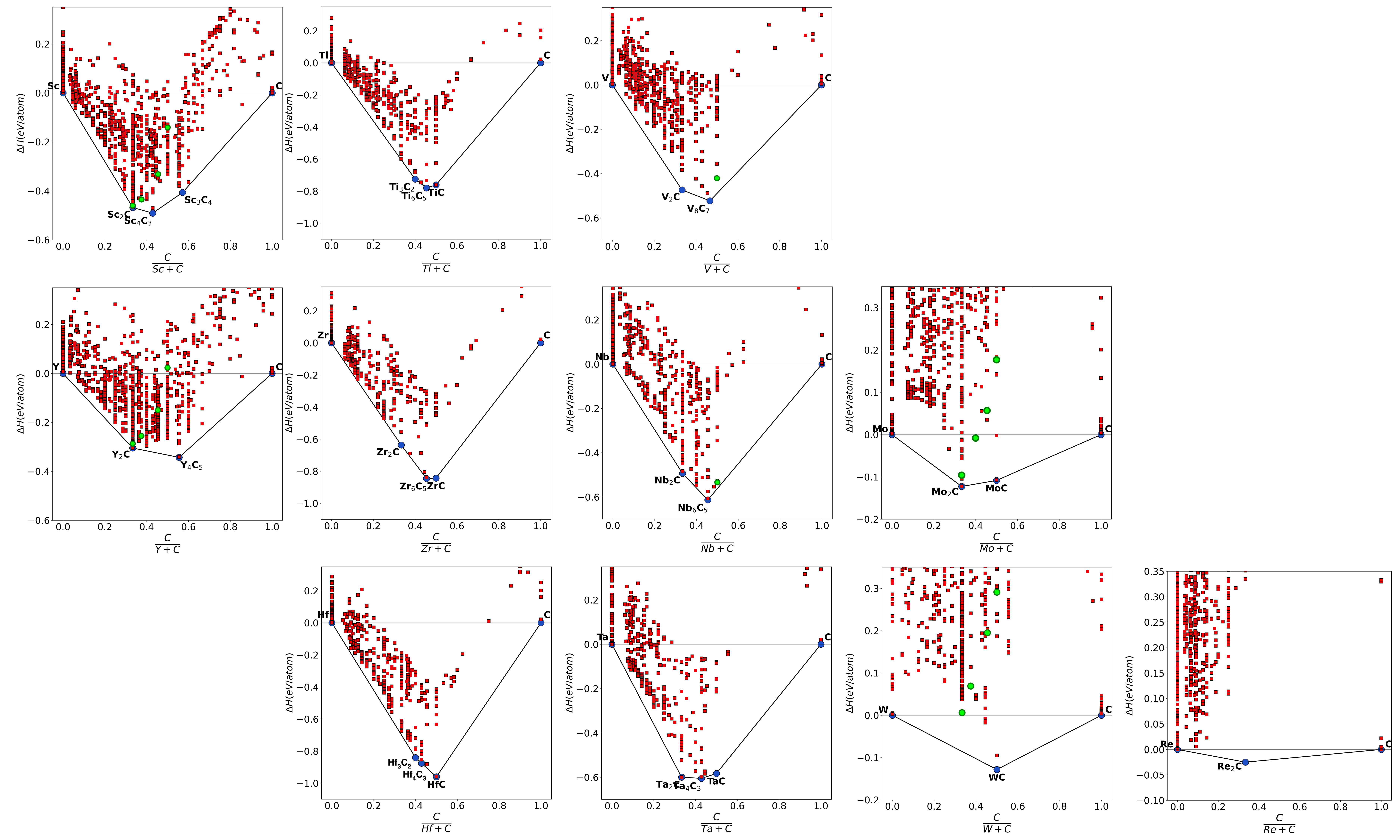} 
    \caption{Convex hulls for TMC's at ambient pressure. The figures are ordered with increasing atomic number, and follow the disposition of the respective TMs in the periodic table. Thermodynamically stable and unstable structures that lie on/above the hull are indicated with blue circles and red squares, respectively. Metastable rock-salt structures with carbon vacancies are marked as green circles when above the convex hull.}
    \label{fig:all_hulls}
\end{figure*}
\end{turnpage}

In Figure~\ref{fig:all_hulls} we show the \emph{ab-initio} convex–hull diagrams 
calculated at ambient pressure for all TMCs in Table \ref{tab:tcs_table}, using evolutionary algorithms for crystal structure prediction \cite{Oganov_JCP_2006_uspex, Lyakhov_cpc_2013_uspex} (See the Methods section and the Supplementary Materials for further details \cite{SM}).

The plots are arranged in order of increasing atomic numbers, from ScC to ReC.
In these diagrams each point indicates the formation energy of one specific structure,
while the lines indicate the convex hull;
points on the hull (blue circles), indicate structures thermodynamically stable against decomposition, whereas structures that lie above the hull are indicated as red squares.
 All computed convex hulls contain at least one stable composition; however, the depth varies systematically across the \(d\)-block: from about 1 eV/atom in group-IV compounds (TiC, ZrC, HfC) to around 0.1 eV in Mo–C, W–C and Re–C. 
 As we will show, this trend mirrors the progressive filling of antibonding \(d\)-states, which are empty in group-IV TMCs, and the consequent weakening of the M–C bond \cite{Hugosson_1999_MoC_theory}. The stoichiometric rock-salt phase is stable for elements of group IV and Ta \cite{note_MoW}. 
However, for elements in groups III, V, and VI, we also find a plethora of \emph{sub-stoichiometric}, carbon-deficient phases—such as M\textsubscript{2}C, M\textsubscript{3}C\textsubscript{2}, and M\textsubscript{6}C\textsubscript{5}—with rock-salt–like geometries (highlighted as green circles in Fig.~\ref{fig:all_hulls}). In several cases, these phases are more thermodynamically stable than the ideal 1:1 compound, or at least weakly metastable.

In many cases, their powder diffraction patterns are nearly indistinguishable from that of the ideal 1:1 NaCl-type structure (see Fig. S2 in the Supplementary Material). This suggests that many TMC samples for which finite critical temperatures were reported in past experiments may have actually contained nonstoichiometric, vacancy-rich phases rather than the nominal 1:1 compound.
Indeed, this hypothesis is strongly supported by the fact that one such structure—M\textsubscript{6}C\textsubscript{5}— is an excellent model for capturing the effect of carbon vacancies in several TMCs.
Before discussing superconducting properties, we first analyze general trends in thermodynamic stability, based on the calculated convex hulls.

\textbf{Group III} (Sc, Y): stable compositions include Y$_2$C and Y$_4$C$_5$; Sc$_2$C, Sc$_4$C$_3$, and Sc$_3$C$_4$, which have been experimentally reported \cite{Giorgi_1968_Y2C, Atoji_1969_Y2C, Gschneidner_BAPD_1986_ScC, Poettgen_1991_Sc3C4, Czekalla_1997_Y4C5, Amano_JPSJ_2004_Y2C3, Juarez_2010_ScC, Juarez_2011_ScC, Aslandukova_PRL_2021_Y4C5}. 
All these the thermodynamically stable structures are qualitatively different from the NaCl structure, as they contain C$_2$, or C$_3$ chains.

The stoichiometric rock-salt structure is over 300 meV/atom above the convex hull. However, as  the concentration of carbon vacancies increases, the relative energy of the rock-salt-like phases decreases drastically, to the point that the vacant rock-salt structure of Sc$_2$C (Y$_2$C) lies just 7 meV/atom (18 meV/atom) above the convex hull.
These carbon-deficient phases may explain experimental reports of superconductivity in Sc and Y 
compounds nominally described as rock-salt carbides.

Experimental reports of a rock-salt ScC phase always involve a rapid quenching from extreme temperatures \cite{Nowotny_1961_ScC, Samsonov_1962_ScC, Juarez_2010_ScC}. As for YC, although its \tc{} is cited in Ref ~\cite{Vonsovsky_1982_TMS}, we were unable to retrieve the original source (Ref. \cite{Samsonov_1961_YC}) to
determine the synthesis conditions.
    
\textbf{Group IV} (Ti, Zr, Hf): the stoichiometric rock-salt structure is always thermodynamically stable. Carbon-vacant phases are present (Ti$_{6}$C$_{5}$, Ti$_{3}$C$_{2}$, Zr$_{6}$C$_{5}$, Hf$_{4}$C$_{3}$, Hf$_{3}$C$_{2}$), and their crystal structures retain a clearly recognizable rock-salt geometry throughout the 2:1 to 1:1 composition range.
These results are in excellent agreement with the experimental report of  synthesis 
of rocksalt structures with a wide range of vacancy concentrations \cite{Klimashin_1970_TiC, Goretzki_1967_TiCZrC}. In addition, for Zr we find on the convex hull a layered hexagonal phase with 2:1 composition, as also predicted in Ref. \cite{Guo_2022_Zr2C}, which has not been experimentally synthesized.

\textbf{Group V} (V, Nb, Ta): the stoichiometric rock-salt phase becomes progressively more stable down the group: the formation enthalpy is 92 meV/atom for V, 28 meV/atom for Nb, and negative for Ta. For all three elements we also find stable carbon-deficient rock-salt phases (V$_{8}$C$_{7}$, Nb$_{6}$C$_{5}$, Ta$_{4}$C$_{3}$) which, in the case of Nb and V, are more stable than the stoichiometric phase.
    
The presence of vacancies in group-V TMCs
has been widely documented experimentally\cite{Henfrey_StrSci_1970_V8C7, Wang_PhysScri_2013_V8C7, Landesman_JPC_1985_Nb6C5, Gao_ChiPhy_2014_NbxCy, Wu_AllComp_2013_NbxCy}. Some authors suggested the formation of vacancy superstructures \cite{Carlson_BAPD_1985_VxCy, Smith_JNM_1987_NbxCy, Kostenko_RuAcSci_2018_M6C5} for both vanadium (V$_6$C$_5$ \cite{Venables_1968_V6C5, Hiraga_PhiMag_1973_V6C5} and V$_8$C$_7$ \cite{Henfrey_StrSci_1970_V8C7, Wang_PhysScri_2013_V8C7}) and niobium (Nb$_6$C$_5$ \cite{Toth_ActMet_1968_NbC_vac, Landesman_JPC_1985_Nb6C5}). 
These observations strongly support  our thermodynamic analysis.

\textbf{Group VI} (Mo, W): the stoichiometric rock-salt phase has a very high formation energy of 288 meV/atom (420 meV/atom) for Mo (W), making it strongly unstable. The thermodynamically stable phase for the 1:1 composition is, in fact a hexagonal phase for both Mo and W,
as confirmed by experiments.\cite{Schuster_1976_MoC_hex}. We also found a 1:2 hexagonal phase on that is on the convex hull for Mo and very close for W \cite{Lautz_1961_W2C_WC, Parthe_1963_Mo2C, Giorgi_1985_W2C, Page_2008_Mo2C}. 
    
Introducing vacancies in the rock-salt phase lattice lowers its relative formation energy dramatically, down to 26 meV/atom in Mo$_2$C (green circles in Fig. \ref{fig:all_hulls}). 
Indeed, the experimental synthesis of the rock-salt phase through high-temperature melting and extreme quenching 
was reported for both MoC and WC \cite{Toth_JPCS_1965_MoC, Morton_JCM_1971_MoC_WC, Willens_AppPhysLett_1965_WC_MoC, Morton_JCM_1971_MoC_WC}. This  resulted in an intermediate, metastable phase stabilized by a certain amount of carbon vacancies \cite{Yvon_1968_W2C}.

\textbf{Group VII} (Re) In rhenium carbide the 1:1 rock-salt phase is 932 meV/atom above the hull, and even carbon vacancies are not enough to stabilize the lattice, with Re$_{6}$C$_{5}$ lying 688 meV/atom above the hull. Recent experiments also failed to reproduce a rock-salt phase \cite{Juarez_2008_ReC}.
We observe instead that  Ref. \cite{Popova_1972_ReC}, which reports a rock-salt phase of ReC, 
reports a lattice parameter of 4.0 \AA. This value is much closer to the calculated equilibrium lattice constant of Re (3.9) than that of ReC (4.4), suggesting that the samples of \cite{Popova_1972_ReC} may have contained mostly pure Re, 
and, possibly, a small concentration of C impurities.

In summary, our results indicate that rock-salt-like TMC phases can be synthesized for a wide range of transition metals,
even in cases where the 1:1 rock-salt structure is thermodynamically or dynamically unstable.
Carbon vacancies play a key role in stabilizing these structures,
lowering their formation energies and making them accessible as metastable phases.

This is reflected in the synthesis conditions reported in the literature: rock-salt–like phases are typically obtained via rapid quenching from the high-temperature melt\cite{Willens_PR_1967_TMC_SC, Nakamura_MSE_2008_TMC_ndiff}.
 Non-equilibrium synthesis methods are consistent with their metastable nature and the presence of a high vacancy concentration. Note that C-vacancies may not be always easy to detect through X-Ray Diffraction analysis, especially if they do not form an ordered superstructure. 

\begin{figure*}[tb]
    \centering
    \includegraphics[width=1.95\columnwidth]{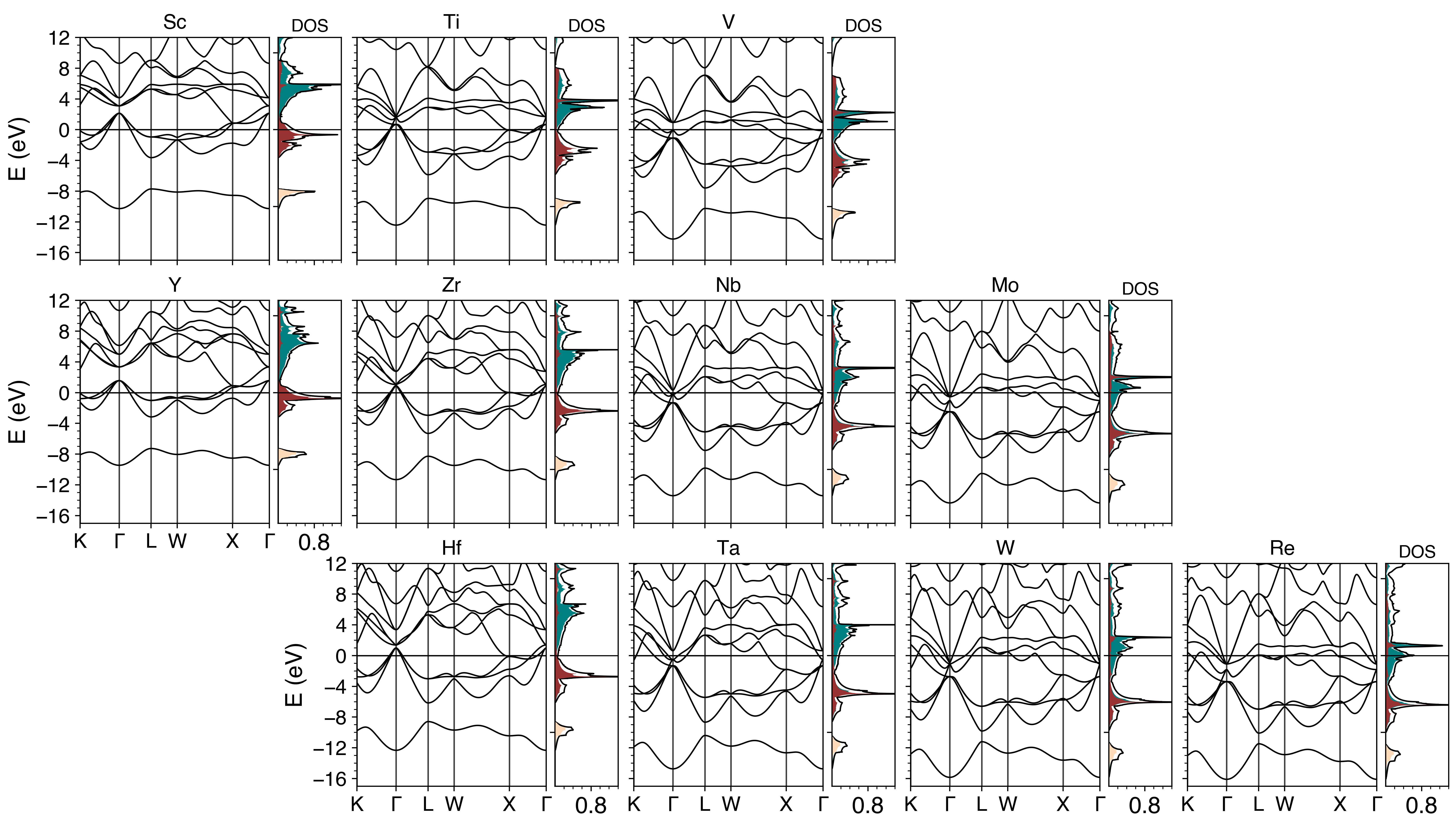}
    \caption{Electronic band structures (left panels) and projected densities of states (right panels) for rock-salt MC (M = Sc–Re), ordered with increasing atomic number, and following the disposition of the respective TMs in the periodic table. Colour code: C–$2s$ (dark red), C–$2p$ (orange) and M–$d$ (teal). The Fermi level is at $(E=0)$. The DOS is in units of states$/$eV$/$spin$/$atom.}
    \label{fig:fig2}
\end{figure*}

\subsection{Electronic structure}
Figure~\ref{fig:fig2} shows the electronic band structures and
projected densities of states (DOS) of the eleven stoichiometric rock-salt carbides, arranged in the same order as the corresponding metals appear in the periodic table. For each TMC the left panel shows the electronic bands, while the right panel displays the DOS decomposed into C–$2s$ (dark red), C–$2p$ (orange), and M–$d$ (teal) contributions. 

The band structures of all compounds comprise the same three groups of bands: (i) a nearly flat, purely C–$2s$ band between -9 and -8 eV; (ii) a broad $\sim 10$-eV-wide manifold around the Fermi level, which consists of six bands; (iii) weakly dispersive, metal-dominated states above. 

The DOS corresponding to the broad manifold around the Fermi level exhibits a characteristic symmetric double feature: a sharp van-Hove peak with C-$2p$ character, which declines towards a pseudo-gap, followed by a rise into another peak, with M-$d$ character.
These six bands derive from the hybridization of the three $d_{t_{2g}}$
orbitals of the M atom with the three C–$p$ orbitals, and are clearly split into three \textit{bonding} and three \textit{antibonding} states, separated by a pseudogap.

The hybrid C–$p$$/$M–$d$ manifold behaves in a rigid-band fashion across the TM series from group III to VII. For ScC and YC the Fermi level sits on the shoulder of the C-$2p$ peak, resulting in a moderate DOS at the Fermi level (listed in Tab. \ref{tab:tcs_table}). In TiC, ZrC and HfC E$_F$ lines up almost
exactly with the pseudogap. For these elements, the TM-C bond strength is maximum,
because the tetravalent M atoms match the preferred carbon valence.

When, in group V (V, Nb, Ta) TMCs, an additional $d$ electron is added, E$_F$ is pushed onto the metal-dominated DOS peak above the gap. Finally, in Mo, W and Re
(group VI) the Fermi level moves closer to the van-Hove singularity, so the DOS is even larger. However, as we shall see, this causes a dynamical instability
of the stoichiometric rock-salt phase.

Moving down a group (e.g.\ Ti\,$\rightarrow$\,Zr\,$\rightarrow$\,Hf) mainly reduces the bandwidth,
since TM with a larger atomic radius form TMCs with larger lattice parameters.
 This effect is particularly relevant in group V,
 where the theoretical equilibrium lattice parameter of Nb and Ta is almost identical (4.47 and 4.48 \AA, respectively),
 but is significantly  larger than that of V (4.15 \AA). As a result, the DOS of V exhibits
 significantly sharper features; in particular, the value at E$_F$ is  50\% larger,
 resulting in a significantly larger \tc{} predicted for the stoichiometric 1:1 phase.
 (See Tab. S3-S4 of the Supplemental Material for the list of calculated lattice parameters \cite{SM}). 

The thermodynamic stability indeed reflects quite closely the filling of the $t_{2g}$-C states:
in group IV, where all bonding states are full, the depth of the hull is largest;
in groups V-VII, as antibonding states are gradually filled, the 1:1 structure
becomes first gradually less stable, and finally unstable in Mo, W and Re.

\begin{figure*}[tb]
    \centering
    \includegraphics[width=1.95\columnwidth]{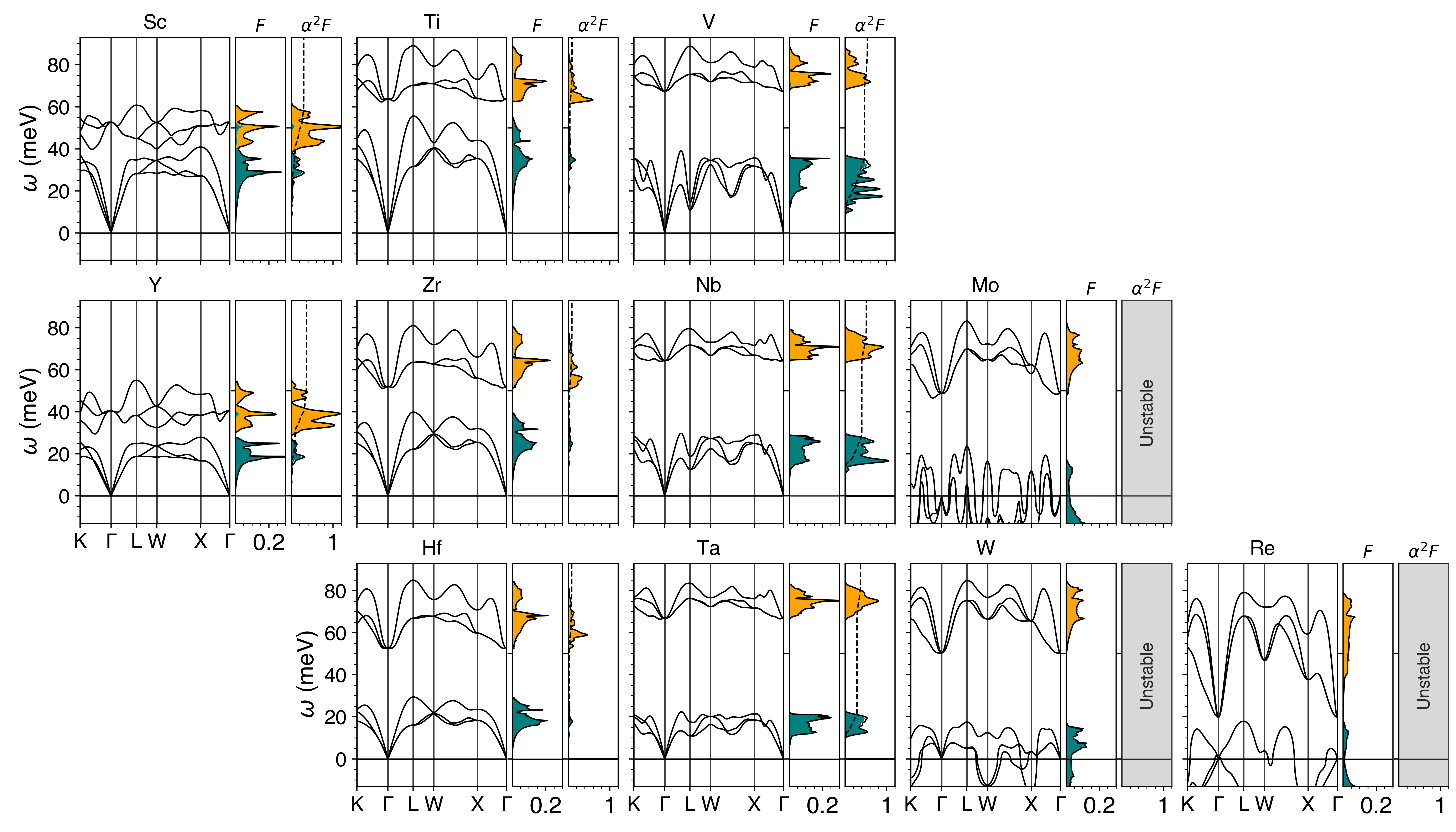}
    \caption{Phonon dispersions, phonon DOS $F(\omega)$ and Eliashberg function $\alpha^{2}F(\omega)$ for rock-salt MC (M = Sc–Re).  Orange (teal) shading indicates the projection onto carbon (transition metal) modes. The dashed line in the rightmost panel is the cumulative integral $\lambda(\omega)$. Grey panels mark dynamically unstable phases, for which we did not compute the electron-phonon properties.}
    \label{fig:fig3}
\end{figure*}

\subsection{Lattice dynamics and electron–phonon coupling}
We computed the phonon and superconducting properties of the TMC series using Density Functional Perturbation Theory \cite{Giannozzi_JPCM_2009_qe, Giannozzi_JPCM_2017_qe} (See the Methods section for further details) in the stoichiometric rock-salt phase for all elements. The crystal structures, as well as further computational details are reported in the Supplementary Materials \cite{SM}. 

Figure ~\ref{fig:fig3} presents the phonon dispersion along the same high-symmetry directions used for the electronic bands, together with the total and atom-projected phonon DOS $F(\omega)$ and the Eliashberg function $\alpha^{2}F(\omega)$.
The panels for different TMCs are
arranged in the same order as Fig. \ref{fig:all_hulls}; the color code
indicates the projection of the eigenmodes on carbon (orange) and transition-metal (teal) vibrations.
The dashed curve in the rightmost panels indicates the frequency-dependent electron-phonon coupling constant $\lambda(\omega)=2\int_{0}^{\omega}\!\alpha^{2}F(\Omega)/\Omega\,
\mathrm d\Omega$.

All spectra exhibit a gap between modes with mainly TM character, 
below $30$ meV, and modes with dominant carbon character, at higher frequencies; the extent
of the spectrum depends on the nature of the TM mode, and range from 60 meV in ScC and YC 
to 85 meV in all other cases.
C-based vibrations are softer in TMCs containing group III elements because the
incomplete filling of the corresponding bonding bands weakens the TM$t_{2g}$-C bonds.
The dispersions of compounds containing  group-IV and group-V elements are quite similar,
although a pronounced softening is clearly seen along the W-X path.
In TMCs containing groups VI and VII elements entire branches become imaginary: the stoichiometric MoC, WC and ReC structures are dynamically unstable. This is consistent both with their high formation energies
and with the calculated electronic structure. In all three compounds,  the Fermi level is very close to a van-Hove peak
in the Density of States, which generally points to an instability towards a lower-symmetry structure with a lower DOS.

The variation in the relative shape of the phonon density of states - $F(\omega)$ - and electron-phonon spectral function - $\alpha^{2}F(\omega)$ - reveal marked differences
in the nature of the electron-phonon
coupling across the series. These differences reflect the same
rigid-band trend observed in the electronic structure.
 In group III TMCs, where electronic states at the Fermi level have essentially C-$p$ character,
 75\% of the total $\lambda$ is concentrated in  carbon vibrations.
In group IV TMCs, where $E_F$ lies in a pseudo-gap of the electronic DOS, the electron-phonon coupling is essentially zero; in group V TMCs, where the electronic DOS at $E_F$ is dominated by TM $d_{t_{2g}}$ states,
75\% of the total  $\lambda$ is due to TM vibrations. 

Overall, the lattice dynamics and the nature of the electron-phonon coupling across the TMC series can be understood in terms of the filling of the metal 
$t_{2g}$–carbon bonding/antibonding manifold:
\begin{itemize}
\item in group III, partial filling of bonding states leads to soft phonons and strong coupling via carbon vibrations;
\item in group IV, full bonding-state filling results in stiff lattices and negligible coupling due to the presence of a pseudogap at $E_F$;
\item in group V, progressive filling of antibonding states enhances coupling through metal vibrations;
\item in groups VI and VII, further filling of the antibonding states ultimately drives dynamical instabilities.
\end{itemize}

\begin{figure}
    \centering
    \includegraphics[width=0.95\columnwidth]{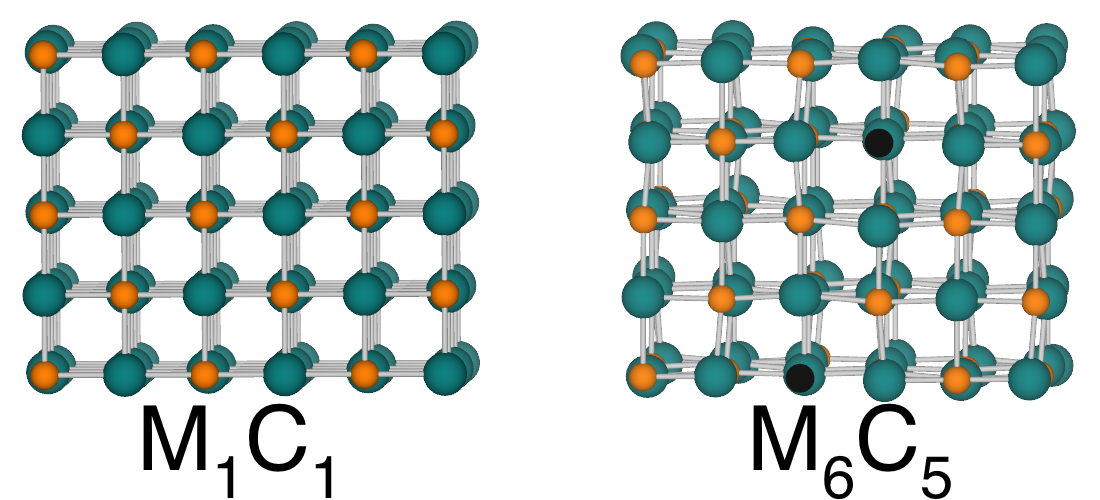}
    \caption{Crystal structures of pristine (M$_1$C$_1$) and carbon-vacant (M$_6$C$_5$) TMCs. The TM, C, and vacancies are indicated as large green, small orange, and small black spheres, respectively.}
    \label{fig:structures}
\end{figure}

\begin{figure*}[htb]
    \centering
    \includegraphics[width=1.95\columnwidth]{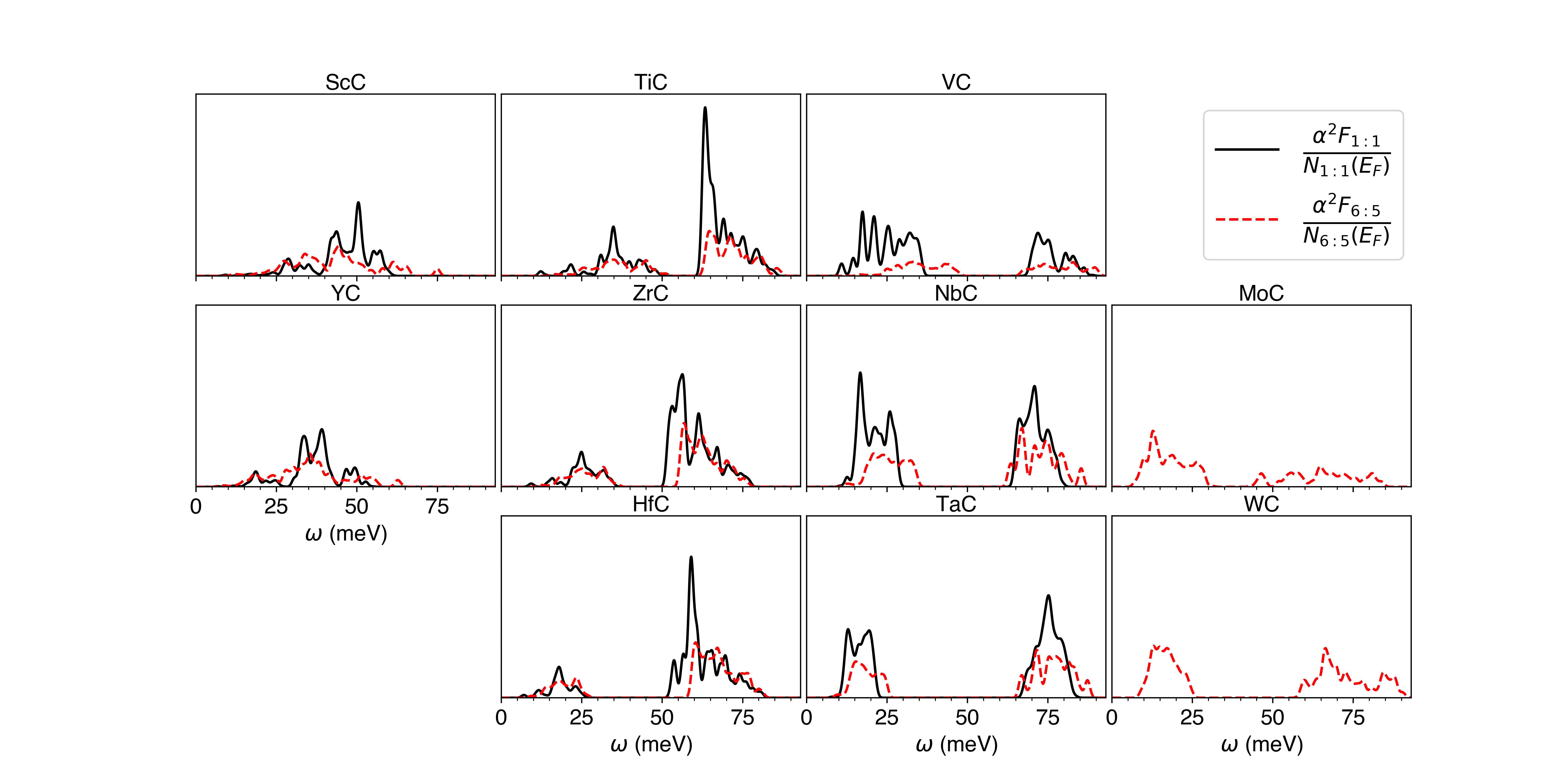} 
    \caption{Comparison of the Eliashberg function ($\alpha^2F(\omega)$ for transition-metal carbides in the rock-salt stoichiometric phase (1:1, black, solid lines) and in the vacant M$_6$C$_5$ phase (6:5, red, dashed lines), rescaled by the electronic DOS at the Fermi energy, across the 3d, 4d, and 5d series.}
    \label{fig:a2f}
\end{figure*}

\subsection{Superconductivity}
Using the calculated $\alpha^2 F(\omega)$ we computed the superconducting \tc{} using the McMillan formula \cite{McMillanTC, Allen_PRB_1975_transition} (See Eq. \ref{eq:mcmillan}), with a a constant value of $\mu^{*}$ of 0.15.

The results are summarized in columns 5-8 of Table \ref{tab:tcs_table}. The \tc{} calculated for the 1:1 stoichiometric rock-salt phase is within 30\% of the experimental value
for only five (Ti, Zr, Nb, Hf, Ta), out of the eleven elements studied, highlighted in green.
For the remaining six, the \tc{} is either largely overestimated (Sc, V, Y), or could not be computed (Mo, W, Re), as the structure is dynamically unstable.

These discrepancies can be understood by analyzing not only the \tc{}, but considering also the enthalpy relative to the hull ($\Delta H$). In all cases for which the predicted \tc{} for the stoichiometric rock-salt phase does not agree with experiment, the formation enthalpy is higher than 50 meV/atom, which is usually considered a practical threshold for metastability; moreover, in Mo, W, and Re, the rock-salt structure is even dynamically unstable. 
In other words, the disagreement does not indicate a limitation of conventional Migdal–Eliashberg theory; rather, it reflects the fact that quantitative predictions of \tc{} become unreliable when based on an incorrect structural model, instead of the true thermodynamic ground-state.

In fact, our analysis raises a crucial point: if the structure if the measured \tc{} does not correspond to the stoichiometric phase — and the calculated values confirm it — then what structure is actually measured in the experiments?

In the previous section, we have shown that carbon-vacant rock-salt-like structures lie on or close to the convex hull across the whole TMC series. In particular, in groups III, V, and VII, they are much more stable than the 1:1 phase. The most plausible explanation is that experiments were performed on such carbon-vacant phases, but the resolution of the powder diffraction spectra was not sufficient to resolve the concentration of carbon vacancies. 

\section{Carbon-vacant phases}
To test this hypothesis, we computed the superconducting \tc{} for a carbon-vacant rock-salt structure with M$_{6}$C$_{5}$ composition (M$_{1}$C$_{0.83}$), shown in Fig. \ref{fig:structures}. 
This structure corresponds to the experimentally observed crystal structure of Nb$_6$C$_5$ \cite{Rempel_1984_Nb6C5},
and was also independently reproduced by our unbiased crystal structure prediction calculations, which systematically placed it on or close to the convex hull across several TMCs.
We selected it as a general, representative example of carbon-deficient rock-salt–like phases.

The specific choice was motivated by the relatively small and symmetric unit cell, which makes \ep{} calculations affordable, and by the weak vacancy–vacancy interactions expected for this geometry \cite{Jang_ActaM_2012_TiC_vacancies}.
 
The  calculated \tc{}'s, reported in the 12th column of Tab. \ref{tab:tcs_table}, are in significantly closer agreement with experiments for all elements that were not already described by the stoichiometric phase, except for rhenium. 
The comparison between the stoichiometric and carbon-vacant phase reveals two main effects: (i) in TMCs containing elements of group III and V, carbon vacancies suppress the \tc{} drastically. In group V, this suppression—combined with the presence of several weakly metastable vacancy-rich phases—naturally explains the broad spread of experimental \tc{} values;
(ii) in group VI compounds, the carbon-vacant structure is dynamically stable, unlike the stoichiometric phase, and the calculated \tc{}'s -- around 18 K for Mo$_{6}$C$_{5}$, and 5 for W$_{6}$C$_{5}$ -- are in good agreement with the reported experimental values. 
In contrast, Re$_{6}$C$_{5}$ is unstable, and no amount of vacancies can stabilize it. It is therefore likely that the \tc{} measured in Ref. \cite{Popova_1972_ReC} originated from pure Re, or some other impurity in the sample.

For group V TMC, the experimental dependence on \tc{} on vacancy concentration has been 
investigated in detail. For VC, the nominal concentration is usually  VC$_{0.88}$, which is not superconducting. Superconductivity can be induced by carbon implantation \cite{Geerk_1977_implantation_VC}, reaching a maximum \tc{} of 3.2 K. In NbC, the dependence of \tc{} on C-vacancy concentration has been studied systematically. 
The rock-salt phase exhibits  a maximum \tc{} of 11.1 K  \cite{Willens_PR_1967_TMC_SC, Toth_2014_TMC}, but superconductivity is strongly suppressed as the 
the number of vacancies is increased \cite{Giorgi_PR_1962_vacancy, Toth_ActMet_1968_NbC_vac, Geerk_1977_implantation_NbC}, vanishing completely for concentrations below $x = 0.77$. Tantalum behaves in a similar way, with a maximum \tc{} of 10.3 K \cite{Toth_2014_TMC}, and a sharp decrease with carbon vacancies \cite{Giorgi_PR_1962_vacancy, Toth_ActMet_1968_NbC_vac}. 

These observations naturally raise the question: why do carbon vacancies suppress superconductivity so strongly in group V TMCs?

The \tc{} of a superconductor can be written in terms of the McMillan formula \cite{McMillanTC}
\begin{equation}
\label{eq:mcmillan}
T_{\mathrm c}
  = \frac{\omega_{\log}}{1.20}\,
    \exp\!\Biggl[
      -\,\frac{1.04\bigl(1+\lambda\bigr)}
              {\lambda
               - \mu^{*}\!\Bigl(1+0.62\,\lambda\Bigr)}
    \Biggr]
\end{equation}
Where $\lambda$, $\omega_{log}$, and $\mu^{*}$ are the \ep{} coupling constant, the logarithmic average phonon frequency, and the Morel-Anderson pseudopotential, respectively. The \ep{} coupling constant can be approximated by Hopfield's expression

\begin{equation}
\label{eq:hopfield}
\lambda=\frac{N(E_F)\,\langle g^{2}\rangle}{M\langle\omega^{2}\rangle},
\end{equation}

A decrease in the superconducting \tc{} is most commonly associated to a decrease in $\lambda$ which, in turn, can originate either from a decrease of $N(E_F)$,  $\langle g^2 \rangle$, or both.

The suppression of $N(E_F)$ is a  well-established mechanism, which is often invoked to
explain the suppression of \tc{}  due to crystal imperfections, as some of us have recently
shown in NbTi. \cite{Cucciari_PRB_2024nbti}.
  
However, this mechanism does not apply to TMCs, as $N(E_F)$ remains essentially unchanged with vacancy concentration in group III and V TMC, while \tc{} is strongly suppressed  -- see columns 5-6 and 11-12 of Tab. \ref{tab:tcs_table} \cite{footDOS}.

Hence, the suppression of \tc{} must  originate from a  reduction of the \ep matrix elements. 
To isolate this effect in Figure \ref{fig:a2f}, we compare the Eliashberg spectral functions of the stoichiometric and carbon-vacant phases, scaled by their respective $N(E_F)$ values. This scaling removes the trivial effect of the DOS in Hopfield's formula -- Eq. \ref{eq:hopfield} -- so that any remaining discrepancy reflects a genuine change in the electron-phonon interaction strength and spectral distribution.

Figure \ref{fig:a2f} shows that in the TMCs of group V the suppression of coupling due to the introduction of carbon vacancies is particularly strong, and is accompanied by a sizable renormalization of the spectrum towards higher frequencies.

\section{Conclusions}

In this work, we reassessed superconductivity in rock-salt transition-metal carbides by combining first-principles electron–phonon calculations with variable-composition evolutionary structure prediction. Despite decades of study and the potential relevance of these compounds for rugged-environment superconductivity applications, a consistent microscopic description is still lacking.  
State-of-the-art electron–phonon calculations show that the commonly assumed 1:1 stoichiometric rock-salt structure fails to reproduce experimental trends in \( T_c \) and is often dynamically or thermodynamically unstable. However, theory and experiment can be reconciled by properly accounting for carbon vacancies. In fact, unbiased evolutionary structure prediction identifies a family of low-energy, rock-salt–like nonstoichiometric phases that are both thermodynamically stable and consistent with the observed superconducting properties. The effect of carbon vacancies on \( T_c \) is highly nontrivial and can only be captured by structurally accurate models.

Our results definitively establish thermodynamic stability as a key factor for the modelling of real-world superconductors. Most large-scale technological applications rely on low-temperature superconductors (LTS) synthesized under strongly non-equilibrium conditions, where defects, disorder, and synthesis history play a central role—yet are typically neglected in state-of-the-art computational techniques.  
By integrating realistic structural thermodynamics into first-principles calculations, our approach sets a practical foundation for predictive modelling of superconductors under realistic synthesis conditions.

\section*{Acknowledgements}
L.B. and S.D.C. acknowledge computational resources from the EuroHPC project "EXCHESS" (EHPC-REG-2024R01-089) and funding from the European Union - NextGenerationEU under the Italian Ministry of University and Research (MUR), “Network 4 Energy Sustainable Transition - NEST” project (MIUR project code PE000021, Concession Degree No. 1561 of October 11, 2022) - CUP C93C22005230007.

\section{Methods}
Evolutionary crystal structure prediction calculations were performed with \textsc{USPEX}~\cite{Oganov_JCP_2006_uspex,Lyakhov_cpc_2013_uspex} using a variable–composition sampling. The local relaxations were performed in a five-step procedure with progressively tighter constraints using \textsc{VASP}~\cite{Kresse_PRB_1993_vasp1,Kress_PRB_1996_vasp2,Kresse_PRB_1999_vasp3}. We employed PAW–PBE potentials, a plane–wave cut-off gradually raised from the recommended minimum up to 500\,eV, Gaussian smearing down to 0.02\,eV, and a $k$–point density down to 0.35\,\AA$^{-1}$.

All electronic, vibrational and superconducting properties were
computed with \textsc{Quantum ESPRESSO} 7.3.1 ~\cite{Giannozzi_JPCM_2009_qe,Giannozzi_JPCM_2017_qe}, after re-optimizing the structures. Optimized norm-conserving Vanderbilt (ONCV) PBE pseudopotentials
\cite{Hamann_PRB_2013_ONCV} were adopted with a 100 Ry cut-off on the plane-wave expansion. A Methfessel–Paxton smearing of 0.02 Ry was used for charge-density integration, and a Gaussian smearing of 0.015 Ry for electron-phonon coupling integration.

Harmonic force constants were obtained within density-functional
perturbation theory and Fourier-interpolated to obtain phonon DOS and dispersions.

Critical temperatures were obtained from the McMillan–Allen–Dynes formula using the calculated \(\lambda\) and \(\omega_{\log}\) and a Coulomb pseudopotential \(\mu^{*}=0.15\). 

Further details are available in the Supplementary Materials \cite{SM}.

\bibliography{library}

\begin{thebibliography}{81}
\bibitem [{not()}]{note_MoW}%
  \BibitemOpen
  \href@noop {} {}\bibinfo {note} {The 1:1 stoichiometry is stable in the MoC
  and WC system, but with an hexagonal structure, while the rock-salt geometry
  is high in energy, as shown in the diagram.}\BibitemShut {Stop}%
\end{thebibliography}

\end{document}